\documentclass{amsart}  
\usepackage[english]{babel}
\usepackage[utf8]{inputenc}
\usepackage{algorithmic}
\usepackage{algorithm}
\usepackage{amsfonts}
\usepackage{amsmath}
\usepackage{amssymb}
\usepackage{amsthm}
\usepackage{enumerate}
\usepackage{epsfig}
\usepackage{graphicx}
\usepackage{amsaddr}
\usepackage{subfigure}
\usepackage{url}

\newcommand{\av}{\textbf{a}}
\newcommand{\bv}{\textbf{b}}

\newcommand{\pv}{\textbf{p}}
\newcommand{\qv}{\textbf{q}}

\newcommand{\xv}{\textbf{x}}

\newcommand{\Sigmam}{\mbox{\boldmath$\Sigma$}}

\newcommand{\diag}{\mbox{diag}}

\newcommand{\Fm}{\textbf{F}}

\newcommand{\Pm}{\textbf{P}}
\newcommand{\Qm}{\textbf{Q}}
\newcommand{\Rm}{\textbf{R}}

\newcommand{\Xm}{\textbf{X}}

\begin{document}

\title{Learning Reputation in an Authorship Network}

\author{Charanpal Dhanjal \and St\'ephan Cl\'{e}men\c{c}on}
\address[Charanpal Dhanjal]{T\'{e}l\'{e}com ParisTech, 46 rue Barrault, 75634 Paris Cedex 13, France}
\email[Charanpal Dhanjal]{\{charanpal.dhanjal, stephan.clemencon\}@telecom-paristech.fr}

\date{\today}

\begin{abstract}
The problem of searching for experts in a given academic field is hugely important in both industry and academia. We study exactly this issue with respect to a database of authors and their publications. The idea is to use Latent Semantic Indexing (LSI) and Latent Dirichlet Allocation (LDA) to perform topic modelling in order to find authors who have worked in a query field. We then construct a coauthorship graph and motivate the use of influence maximisation and a variety of graph centrality measures to obtain a ranked list of experts. The ranked lists are further improved using a Markov Chain-based rank aggregation approach. The complete method is readily scalable to large datasets. To demonstrate the efficacy of the approach we report on an extensive set of computational simulations using the Arnetminer dataset. An improvement in mean average precision is demonstrated over the baseline case of simply using the order of authors found by the topic models.  
\end{abstract}

\keywords{reputation assessment, expert finding, graph centrality, rank aggregation}

\maketitle

\section{Introduction}

Identifying experts is a valuable task for finding coauthors for a new research project or grant, assigning reviewers for the peer-review of an article or employing consultants. In so-called Reputation Systems \cite{resnick2000reputation} one has explicit ratings of reputation such seller feedback provided on the eBay online auction site. Here we address the more challenging problem of estimating the reputation of authors in a network of authors and their publications. In a general sense, one must first evaluate the domain(s) of authors and then grade their expertise by the number and quality of publications in peer-reviewed journal and conferences. 

The particular problem under study is stated in a more formal setting as follows.  An undirected graph $G = (V, E)$ is composed of vertices $\{v_1, \ldots, v_n\} = V$ and edges $E \subseteq V \times V$ in which vertices represent authors and edges represent connections between the authors, for example common mediums of influence such as coauthorship or citation. Each vertex has a list of articles associated with it, representing an author's publications. The first question is how can one find all authors who have worked in a given domain $D_i \subset V$ based on their publications. Next, consider a class of scoring functions over the vertices in $D_i$, $f \in \mathcal{F}$, and an unordered set of top $k$ vertices $S_i = \{v_{x_1}, u_{v_2}, \ldots u_{v_k}\} \subset D_i$. Our task is to find a ranking function close to the \emph{oracle} $f^*$ for which the top $k$ ranked elements are identical to $S_i$.  Thus, the learning problem is to identify the characteristics of a reputable author in domain $D_i$ in the space of functions $\mathcal{F}$. 

\begin{figure}[htp]
\begin{center}
	\includegraphics[scale=0.4]{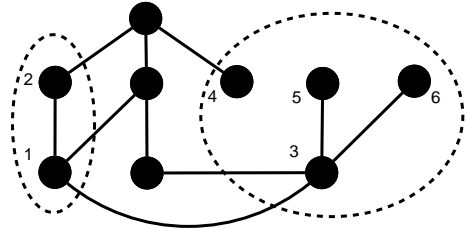}
\end{center}
\caption{A graph of authors and connections, with the authors in the domain of interest circled. If we take the number of edges incident to a vertex as a measure of its expertise then we can see that author $3$ has the most expertise as he/she has 3 edges to other authors in the same domain.}
\label{fig:exGraph}
\end{figure}

To tackle this problem we first study the titles and abstracts corresponding to the articles written by each author using Latent Semantic Indexing (LSI, \cite{deerwester1990indexing}) and Latent Dirchlet Allocation (LDA, \cite{blei2003latent}). These are two popular and effective topic modelling algorithms, and readily scalable to the large dataset typically in use. This step identifies authors within the field of interest. We then construct a coauthorship graph using the authors found in this step (see Figure \ref{fig:exGraph}), and use a variety of centrality measures to score and rank vertices in that particular domain. Furthermore, to leverage the rankings we examine rank aggregation for the task of expert prediction. The chief novelty of this paper is the use of efficient topic modelling approaches and state-of-the-art graph-based algorithms in combination with rank aggregation to study this problem. 

We start by describing our approach to the domain identification of the authors in Section \ref{sec:domain}. Following, we outline a number of centrality measures in graphs and how the ranking of vertices of these measures can be aggregated in Sections \ref{sec:ranking} and \ref{sec:rankAggregation} respectively. Section \ref{sec:related} reviews related work in this area, and then we present computational results on a large author-publication dataset in Section \ref{sec:simulations}. A summary and some perspectives are given in the final section.   

\section{Domain Identification}\label{sec:domain} 

It is a common problem in information retrieval to recover documents corresponding to a particular subject and here we briefly review LSI and LDA and their online variants for this purpose. 

Imagine that one has a set $T = \{d_1, \ldots, d_m\}$ of documents and we wish to discover the subset of those documents in a particular domain. The first step is to preprocess the words using a Porter Stemmer \cite{van1980new} to amalgamate words with the same base such as ``learning'' and ``learned''. One then finds a \emph{bag of $\ell$-grams} representation of the documents which is essentially a count of each sequence of $\ell$ consecutive words in the documents. This enables the identification of important word concurrences such as ``singular values'' which would be lost if a bag of words ($1$-gram) representation was used. This representation can be improved by removing stop words such as ``and'' and ``the'' which are common and do not convey information useful for discrimination.  At the end of this process we have a set of terms ($n$-grams) and documents, and occurrences of terms within the documents. 

Rather than use this data directly it is often useful to  represent documents using the \emph{term frequency/inverse document frequency} (TF-IDF, \cite{salton1989automatic}) representation.  Assume that there are $m$ documents and $k_j$ appears in $n_i$ of them so that $\Fm_{ij}$ is the number of times $k_i$ appears in document $d_j$. The normalised term frequency is defined as: 

\begin{displaymath} 
 \mbox{TF}_{i,j} = \frac{\Fm_{ij}}{\max_z \Fm_{zj}}, 
\end{displaymath}
\nolinebreak
where the maximum is computed over all frequencies $\Fm_{zj}$ for document $d_j$. Frequent keywords may not be useful and hence one also uses inverse document frequency, defined for a keyword $k_i$ as 
\begin{displaymath}
\mbox{IDF}_i = \log \frac{|T|}{n_i}. 
\end{displaymath}
\nolinebreak
The TF-IDF weight for a keyword $k_i$ in document $d_i$ is then $\Xm_{ji} = \mbox{TF}_{i,j} \times \mbox{IDF}_i$ and each document can be represented using a vector of keyword weights. In this way similar vectors correspond to similar documents. 

In LSI, a partial Singular Value Decomposition (SVD, \cite{golub2012matrix}) is performed on the TF-IDF matrix $\Xm$ to determine relationships between the terms and semantic concepts represented in the text. The SVD of $\Xm \in \mathbb{R}^{m \times \ell}$ is the decomposition 
\begin{displaymath}
\Xm=\Pm \Sigmam \Qm^T,
\end{displaymath}
where $\Pm = [\pv_1, \ldots, \pv_r]$, $\Qm = [\qv_1, \ldots, \qv_r]$  are respective matrices whose columns are left and right singular vectors, and $\Sigmam = \diag(\sigma_1, \ldots, \sigma_r)$ is a diagonal matrix of singular values $\sigma_1 \geq \sigma_2 \geq, \ldots, \geq \sigma_r$, with $r = \min(m, n)$. The matrix $\Pm$ can be thought of a mapping from documents to a semantic concept, $\Qm$ is a mapping from terms to concepts and $\Sigmam$ is a scaling of the columns and rows respectively of these matrices. By taking the partial SVD one chooses the singular values and vectors corresponding to the largest $k$ singular values, denoted by $\Pm_k, \Sigmam_k$ and $\Qm_k$ respectively. This truncation has the effect of retaining the important concepts whilst removing noise in the concept space of $\Xm$. Notice that $\Xm$ is typically a sparse matrix and hence one can use efficient methods for computing the SVD such as Lanczos or Arnoldi (e.g. PROPACK \cite{larsen1998lanczos}) or randomised methods  \cite{halko2011finding}. In the later experiments we use the multipass stochastic online LSI algorithm presented in \cite{vrehuuvrek2011subspace}.

LDA is a generative model that explains a set of documents using a small set of topics. It assumes a set of $k$ topics about the set of documents $T$. Each topic is drawn from a Dirichlet distribution $\beta_\ell \in \mbox{Dirichlet}(\eta)$. For each document $d_j$ one draws a distribution over topics $\theta_{d_j} \in \mbox{Dirichlet}(\alpha)$. For each word $t_i$ in the document one draws a topic index $z_{d_j, t_i} \in \{1, \ldots, k\}$ with weights $z_{d_j, t_i} \in \theta_{d_j}$. The observed word is then drawn from $t_{ji} \in \beta_{z_{d_j, t_i}}$. To infer the distributions in this model, one uses a variational Bayes approximation of the posterior distribution. In our later computational work, we use the online variant of LDA given in \cite{hoffman2010online}.  

One approach to evaluate the similarity of a query to the training documents is to map the query and documents to the LSI or LDA space and find the highest cosine of the angle between them, known as \emph{cosine similarity}. Note that the cosine of the angle between two vectors $\av$ and $\bv$ is given by $\cos(\theta) = \av^T\bv/(\|\av\| \|\bv\|)$. If we fix a threshold $\gamma$ and find all documents with $\cos(\theta) > \gamma$ and then the corresponding authors, we have two effective methods of identifying authors in the query domain.  

\section{Ranking Experts} \label{sec:ranking}

Once we have a collection of authors who have published under a particular domain, we can extract the corresponding coauthorship graph and use this graph to rank authors by their expertise. To do so we draw upon state-of-the-art results in graph structure analysis and rank aggregation. The key idea is to construct $\mathcal{F}$ in such a way that we encapsulate the main characteristics of reputation. For that reason, we consider six measures: \emph{influence maximisation} \cite{kempe03maximising}, \emph{PageRank}, \emph{hub score}, \emph{closeness centrality}, \emph{degrees}, and \emph{betweenness}, and motivate their use.  

\subsection{Influence Maximisation}

Influence maximisation is an intuitive way to measure the reputation in an authorship graph. To find the most influential vertices we first introduce the concept of \emph{graph percolation} in which vertices within a graph have a binary state: either active or inactive. A percolation process decides how activation spreads within the graph. The problem of influence maximisation is to find the $k$ vertices which result in the largest total spread of activation at the end of the process\footnote{A percolation process can be said to be concluded when there are no additional activations/disactivations.}. In epidemic spread, for example, finding the most influential vertices may help to devise effective control strategies. 

\begin{figure}[htp]
\begin{center}
	\includegraphics[scale=0.2]{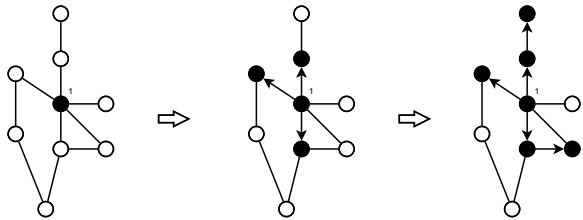}
\end{center}
\caption{Percolation  within a graph. Vertices in black are active and activation spreads to other vertices in an iterative manner. The initial active set is $A = \{1\}$ and $\sigma_P(G, A) = 6$ in this case.}
\label{fig:percolation}
\end{figure}

A binary percolation process $P$ computes in an iterative manner which vertices will be active in the next iteration based on the edges and those that are currently active, and continues until no more activations occur. Let $\sigma_P(G, A)$ be the number of active vertices at the end of a percolation process defined by $P$, over graph $G$ and with an initial set of active vertices $A$. Figure \ref{fig:percolation} demonstrates a percolation process within a simple graph. A commonly studied percolation process is the \emph{Independent Cascade model}. For this model, there is a probability $p_{ij}$ on an edge from $v_i$ to $v_j$ which allows a random decision to be taken for the activity of $v_j$ given that $v_i$ is active. The percolation then proceeds as follows: at time step $t$ when a vertex $v_i$ first becomes active it is given a single chance to activate each of its neighbours $n(v_i)$  according to the edge probabilities. If $v_i$ succeeds then the corresponding vertices become active in the next time step. If not then no further attempts are made in subsequent rounds. 

The problem of computing the maximal influence is: given a graph $G$ and process $P$ which subset $A \subseteq V$ of $k = |A| < n$ vertices should be chosen to ensure maximal activation $\sigma_P(G, A)$ at the end of the process? This optimisation corresponds to a combinatorial problem that is NP-hard in general. The authors of \cite{kempe03maximising} tackle this problem for the Linear Threshold and the Independent Cascade models, and provide greedy algorithms to maximise influence. The key observation is that the influence function $\sigma_P(G, A)$ is \emph{submodular} for these percolation models. A submodular function is one in which there are diminishing returns. Provided that $\sigma_P(G, A)$ is submodular and monotone, a simple $O(nk)$ greedy algorithm exists for choosing the most influential vertices, see \cite{kempe03maximising} for more details. A faster variant of this algorithm, known as Cost Effective Lazy Forward selection (CELF), is given in \cite{leskovec2007cost} in which computational savings are made by using the submodularity property and the previous influences of each vertex at each stage.

\subsection{Graph Centrality Measures}

Closely related to the influence of vertices within a graph is the idea of graph centrality, and here we outline several useful measures.  We begin with PageRank which was designed to rank Web pages using the graph of hyperlinks, however since then has been applied to many other types of graph.

The key intuition of PageRank is that a hyperlink to a page counts as a vote of support, and hyperlinks from ``important'' pages are weighted higher than unimportant ones. In this sense Pagerank is defined recursively and depends on the  PageRank metric of all pages that link to it (incoming links). A page that is linked to by many pages with high PageRank receives a high rank itself. The idea is linked closely to the concept of performing a random walk on a graph. For a directed graph $G$ PageRank is defined as follows: 

\begin{displaymath} 
P(u) = \frac{1-|V|}{D} + D \sum_{n_{in}(u)} \frac{P(v)}{|n_{out}(v)|},
\end{displaymath}
\noindent
where $n_{in}(u)$ is the set of all vertices with edges directed towards $u$, $n_{out}(u)$ is a set of vertices with edges directed from $u$ and $D$ is a damping factor between $0$ and $1$ which is used to enable the random walker to jump out of cycles. 

A precursor to PageRank is the hub score (HS, \cite{kleinberg1999hubs}) of the vertices in a graph. In short, a hub is a catalogue of information that points to authority pages. A highly rated hub points to many authority pages, and a good authority is referenced by many hubs. To compute the hub score we initialise two scores $h(v) = 1$ and $a(v) = 1$ for all $v \in V$. To update these scores one performs mutual recursion as follows: 

\begin{displaymath} 
 a(v) = \sum_{u \in n_{in}(v)} h(u),
\end{displaymath}

and 

\begin{displaymath} 
 h(v) = \sum_{u \in n_{out}(v)} a(u),
\end{displaymath}
and one normalises uses the 2-norm of the corresponding scores after each iteration to allow convergence. 

Another useful measure which we shall use for our analysis is betweenness, which is the number of times a shortest path passed through a certain vertex. Intuitively, this quantifies the importance of the vertex in terms of linking other vertices. It is defined more formally as 

\begin{displaymath} 
B(u) = \sum_{v, w \in V\backslash u} \frac{\sigma_{vw}(u)}{\sigma_{vw}},
\end{displaymath}
\noindent
where $\sigma_{vw}(u)$ is the number of shortest paths from $v$ to $w$ that pass through $u$ and $\sigma_{vw}$ is the total number of paths between $v$ and $w$. 

Next we look at closeness centrality \cite{freeman1979centrality} which is a measure of how close all other vertices are to the current one. One way of considering this type of centrality is how long it would take for information to spread to other vertices in the network, and hence it makes sense for the type of network considered. The closeness centrality is defined as follows: 

\begin{displaymath} 
C(u) = \frac{1}{\sum_{v \in V \backslash u} d(u, v)},  
\end{displaymath}
\noindent
where $d(u, v)$ is the distance between $u$ and $v$. Hence closeness centrality is the inverse of the average length of the shortest paths to all other vertices in the graph. 

Finally we also use the degree of vertices as a measure of their centrality where the degree is simply the number of edges incident to a vertex. 

\section{Aggregating Rankings}\label{sec:rankAggregation} 

In this section we show how to combine the rankings given by the above centrality measures. Rank aggregation has been studied using Borda count \cite{aslam2001models}, median rank aggregation \cite{fagin2003efficient} and Markov Chains \cite{dwork2001rank}. Here we detail the popular Markov chain method of \cite{dwork2001rank}. The principal advantages of Markov Chain based rank aggregation methods is that they can work with partial lists, are efficient, and shown to outperform other methods in \cite{renda2003web}. 

The setup for rank aggregation is described as follows. Consider a set of elements $D$ and an ordered list $\tau$ whose elements are a subset of the elements of $D$, $\tau = [x_1 \geq x_2 \geq \ldots \geq x_{|\tau|}]$ with $x_i \in D$, where $\geq$ is an ordering relation on $D$. If $\tau$ contains all the elements in $D$ it is called a \emph{full list} otherwise it is a \emph{partial} or \emph{top}-$k$ \emph{list} for which only the first $k$ elements are present. In the case of rank aggregation we have a number of ranked lists $\tau_1, \ldots \tau_n$ and we also have an ideal ranking $\tau^*$. The goal is to find an aggregation function $\phi : \tau_1, \ldots, \tau_\ell \mapsto \xv$, where $\xv$ is a score vector for all entries, such that the ordering according to $\xv$ is as close to $\tau^*$ as possible. 

In the $\mbox{MC}_2$ model of \cite{dwork2001rank} we construct a Markov chain which is a state transition machine in which a transition to a new state is dependent only on the current one. Each item $x_i \in D$ is represented by a state and then a ranking list $\tau_j$ is selected randomly such that $x_i$ is an element of $\tau_j$. One then selects a random state uniformly from the elements in $\tau_j$ which are not ranked lower than $x_i$. More formally, define the $k$th \emph{transition matrix} as $\Pm^{(k)}$ such that $\Pm^{(k)}_{ij}$ is the conditional probability of state $x_j$ given state $x_i$ and ranking list $\tau_k$. We have 
\begin{displaymath} 
 \Pm^{(k)}_{ij} = \left\{ \begin{array}{l l} \frac{1}{q} & x_j \geq x_i\\ 0 & \mbox{otherwise},    \end{array}\right.
\end{displaymath}
where $q = |\{x_j | x_j \geq_{\tau_k} x_i\}|$. The final transition matrix is given by the mean of the individual matrices for each ranked list, $\Rm = \frac{1}{\ell}\sum_{i=1}^\ell \Pm^{(i)}$. The score vector is then computed as the stationary distribution $\xv = \Rm^T\xv$ such that $\sum_{i=1}^{|D|} \xv_i = 1$ and $\xv_i > 0$ for $i=1, \ldots, |D|$. 

\section{Related Work}\label{sec:related}

A key driver of expert recommendation in recent years has been the expert finding task in the TREC Enterprise track in 2005 \cite{craswell2005overview}. The data present in this task includes email on public mailing lists, code and web pages extracted from the World Wide Web Consortium (W3C) sites in June 2004. The task consisted of ranking 1092 experts from the 331,037 documents available. Two types of model appeared from this task \cite{baileyoverview2007, balogoverview2008}: candidate and document models. In candidate models, one builds a textual representation of the experts and ranks them based on a query. In document based models, one first finds documents relevant to the query and then locates associated experts. In our work we use a mixture of these two ideas.  

For academic networks, the topic of discovering experts using graphs has been studied in conjunction with the Arnetminer \cite{tang2007arnetminer} academic database and social network in \cite{tang2009social}. Unlike our work which incorporates topic learning as part of the process, the authors label each member of the social network with a pre-assigned topic vector, and then try to measure influence in the network. A Topical Affinity Propagation (TAP) model is proposed which optimises the topic-level social influence on a network. 

In  \cite{deng2008formal} which focuses on the expert seeking task on the Digital Bibliography and Library Project (DBLP) dataset, three models are proposed, namely a Bayesian statistical language model, a topic-based model and a hybrid one. One of the key parts of the model is computing prior probabilities of authors using citation data which is used in conjunction with the language-based ranking of authors. Note that each article is augmented with similar documents extracted from Google Scholar which we do not use in our experiments. The experts are manually graded on a scale from $0-3$ and the learning system is tested against these ratings with favourable results to related algorithms in \cite{li2007eos, zhang2007expert}. In \cite{li2007eos}, the authors augment the DBLP data with Google search results as well as publication rankings from Citeseer. 

A more scientometric approach is given in \cite{heck2011expert} which uses measures such as bibliographic coupling (two authors A1 and A2 are linked if they cite the same references) and author cocitation to recommend similar authors. In this paper there is not a focus on finding the most influential authors. Along the same lines, there are a number of other metrics one could use to rate the reputation of authors in a particular domain such as $h$-index and impact factor. The $h$-index is the largest number $h$ such that $h$ publications have at least $h$ citations and impact factor\footnote{Note that although this is the generally accepted definition of impact factor, alternative definitions exist.} of a journal is the average number of citations in the two preceding years. 

\section{Simulations}\label{sec:simulations}

In this section we evaluate the expertise ranking algorithms by comparing them to the baseline case of using the author order given solely using topic modelling and not any graph-based ranking scheme. The Arnetminer dataset \cite{tang2008arnetminer} which is based on DBLP, is used. This dataset is a list of articles in computer science, along with their authors, the publication venue, year and paper abstracts and citations for some articles. We use version 5 of the dataset which contains 1,572,277 papers with 529,499 abstracts and is generated on 21/2/2011. We want to observe the accuracy of our expert finding approach on this dataset in conjunction with experts in 13 fields suggested on the Arnetminer web site. The expert lists are generated using the Program Committee members of well know conferences/workshops and the members on sites specific to a particular field, for example on \url{www.boosting.org}. The lists are unordered and ``noisy'' due to their nature, however still useful for the purposes of evaluation. We use the experts in the fields listed in Table \ref{tab:dblpExperts}. All experimental code is written in Python and we use the Gensim library \cite{rehurek_lrec} for topic modelling. 

Before predicting a set of experts, we perform model selection for our learning algorithm and hence split the experts into a 50:50 training/test set.  The word vectoriser is set up as described above on the title and abstracts of articles for 1 and 2-grams with term counts included if the term frequency is in at least a proportion $\rho \in \{10^{-3}, 10^{-4}\}$ of the total number of documents. Since the documents being processed are typically small we use binary indicators for terms. For LSI we take the SVD of this matrix using the randomised SVD method of \cite{halko2011finding} with an exponent of $q=2$ and oversampling of $p=100$ and take $k \in \{100, 200, \ldots, 600\}$. After this stage, we find similar documents to a query term (the field) using the method outlined above and a cosine similarity threshold of $\gamma \in \{0.0, 0.1, \ldots, 0.9\}$. Each author in this set is then scored by summing the cosine similarity of their articles and we take the first $x$ authors according to their score (denote this set of authors as $U$). In this case $x$ is 10 times the number of training experts. For LDA we choose the number of topics in $k \in \{100, 200, \ldots, 600\}$ and otherwise use an identical process. The optimal model is selected by choosing parameters which result in the largest number of training experts across the complete set of 13 fields. 
\begin{table} 
\begin{center}
\begin{tabular}{l l l }
\hline 
Abbreviation & Category & Experts  \\  
\hline 
BS & Boosting & 57 \\ 
CV & Computer Vision & 215 \\ 
CRY & Cryptography & 174 \\
DM & Data Mining & 351 \\ 
IE & Information Extraction & 91\\ 
IA & Intelligent Agents & 30 \\  
ML & Machine Learning & 76 \\ 
NLP & Natural Language Processing & 54 \\ 
NN & Neural Networks & 122 \\
OA & Ontology Alignment & 56  \\ 
PL & Planning & 26 \\ 
SW & Semantic Web & 412 \\ 
SVM & Support Vector Machines & 111 \\ 
\hline 
\end{tabular}
\end{center}
\caption{Summary of the information about experts over the Arnetminer dataset.}
\label{tab:dblpExperts}
\end{table} 

After model selection, the authors in $U$ are positioned in a coauthorship graph in which an edge exists only if two authors $u, v \in U$ have collaborated. Edges in this graph are weighted according to the number of articles written by the corresponding pair of authors. We compute each centrality metric over the weighted graph and use the inverse of the weights for the computation of betweenness and closeness centrality. This implies for example that two authors who have collaborated 5 times have an edge weight between them of $1/5$ and thus are more likely to be on a shortest path than adjacent authors who have collaborated less frequently. For influence maximisation we obtain the 100 most influential authors using 100 repetition of the independent cascade model with transition probability $p=0.05$. We also record the order of authors given by the topic modelling approaches and that given by sorting authors according to the total number of citations for articles in $U$.  

The rankings are evaluated using the test set with the Mean Average Precision (MAP) at $N$ metric. The \emph{precision} at $N$ is the number of experts in the first $N$ items of the ranked list of authors divided by $N$ or equivalently
\begin{displaymath} 
p@N = \frac{tp}{tp+fp}, 
\end{displaymath}
where $tp$ is the number of true positives and $fp$ is the number of false positives. Precision falls within the range $[0, 1]$ with 1 signifying that all items at the top of the list are experts.  The \emph{average precision} is the average of all precisions for all of the experts: 
\begin{displaymath} 
ap@N = \frac{\sum_{i=1}^N p@i \times \mbox{rel}(i)}{R}, 
\end{displaymath}
where $\mbox{rel}(i)$ is an indicator function which is 1 for relevant experts and $R$ is the number of experts. MAP is simply the average precision over all the queries. We look at MAP for $N \in \{5, 10, \ldots, 50\}$. Note that to compute these precisions for the test experts we remove the training experts from the rankings, and vice versa. After computing the graph and topic-based rankings, we use the MC2 algorithm of \cite{dwork2001rank} to aggregate rankings from each field in a greedy fashion: using the training experts we pick the ranking with the best $ap@20$ score then choose additional rankings that give the best marginal gain until no improvement is obtained. 

The proportion of training experts covered by LSA topic modelling is approximately 0.381 using $k=500$, $\rho=10^{-4}$ and $\gamma=0.3$. This indicates the difficulty of finding relevant authors using this dataset. It is worth noting that amongst the complete set of experts, a mean proportion of 0.4 of their articles also have abstracts and we believe results could be improved with a higher proportion of abstracts. LDA was less effective than LSI at recovering the training experts during model selection with a mean coverage of 0.318 over all the fields, using $k=400$, $\rho=10^{-3}$ and $\gamma=0.4$.

Table \ref{tab:lsiMap} shows the MAP values on the test experts in conjunction with the authors returned using LSI. In this table we see that the strongest single method is betweenness followed by the citation and topic orders. A possible reason for the efficacy of betweenness is that reputable authors are also social and attract collaborations and hence participate in many shortest paths in the coauthorship graph. Citation is a good indication of reputation since citations are often positive votes about the value and quality of a paper. We observed that the relative performances of the rankings varied between fields. In Ontology Alignment for example the $ap@50$ score for the citation ranking was 0.08 versus 0.163 for betweenness. A particularly challenging field was Neural Networks which was ranked best using influence with $ap@5 = 0.05$ and $ap@50=0.058$. A possible reason is that Neural Networks covers a large range of topics both in biology and machine learning. In contrast, we obtained $ap@5 = 0.76$ with Data Mining using betweenness and $ap@5 = 0.76$ for Information Extraction using the citation ranking. A significant improvement is gained by aggregating the rankings of the topic modelling order, citation ranking and betweenness ranking. We see that $ap@5$ improves from 0.196 using betweenness to 0.250 with $\mbox{MC}_2$ and $ap@50$ improves from 0.152 with citations to 0.168. 

Table \ref{tab:ldaMap} shows the corresponding results using LDA for topic modelling. The best performing rank methods were closeness and PageRank with $ap@5$ scores of 0.137 and 0.134 respectively. They improve significantly over the baseline topic order score (denoted ``Topic'' in the table) of 0.111. Interestingly, the citation-based ranking does not perform well in this case because more irrelevant, but highly cited, authors are found in $U$ relative to LSI. The rank aggregate of our greedy $\mbox{MC}_2$ algorithm gives a slight improvement over the using closeness centrality. When considering individual fields, the comparison to LSI is more complicated. In the case of Semantic Web for example, PageRank gives $ap@5$ of 0.483 using LSI and 0.8 using LDA. As with LSI however, LDA scores poorly when the domain is Neural Networks. 

\begin{table*}
\centering
\begin{tabular}{l | l l l l l l l l l}
\hline
N & Topic & Cit. & Bet. & Cls & PR & Dgr & Inf. & HS & $\mbox{MC}_2$\\
\hline 
5 & 0.152 & 0.170 & 0.196 & 0.130 & 0.129 & 0.065 & 0.112 & 0.000 & 0.250\\
10 & 0.109 & 0.133 & 0.122 & 0.094 & 0.098 & 0.046 & 0.077 & 0.004 & 0.176\\
15 & 0.092 & 0.138 & 0.108 & 0.095 & 0.090 & 0.038 & 0.075 & 0.007 & 0.165\\
20 & 0.087 & 0.141 & 0.110 & 0.092 & 0.086 & 0.036 & 0.077 & 0.008 & 0.161\\
25 & 0.088 & 0.139 & 0.107 & 0.093 & 0.084 & 0.033 & 0.080 & 0.009 & 0.157\\
30 & 0.087 & 0.143 & 0.109 & 0.092 & 0.086 & 0.032 & 0.083 & 0.013 & 0.162\\
35 & 0.086 & 0.146 & 0.118 & 0.096 & 0.085 & 0.033 & 0.085 & 0.013 & 0.165\\
40 & 0.088 & 0.147 & 0.122 & 0.098 & 0.086 & 0.033 & 0.085 & 0.018 & 0.163\\
45 & 0.087 & 0.150 & 0.124 & 0.098 & 0.088 & 0.034 & 0.085 & 0.019 & 0.166\\
50 & 0.089 & 0.152 & 0.126 & 0.099 & 0.089 & 0.034 & 0.085 & 0.022 & 0.168\\
\hline
\end{tabular}
\caption{MAP values at each value of $N$ for the rankings using LSI for topic modelling.  Abbreviations: Topic (LDA order), Cit. (citation order), Bet. (betweenness), Cls (closeness), PR (PageRank), Dgr (degree), Inf. (influence), HS (hub score) and $\mbox{MC}_2$ is the Markov chain model of \cite{dwork2001rank}.}
\label{tab:lsiMap}
\end{table*}

\begin{table*}
\centering
\begin{tabular}{l | l l l l l l l l l}
\hline
N & Topic & Cit. & Bet. & Cls & PR & Dgr & Inf. & HS & $\mbox{MC}_2$\\
\hline 
5 & 0.111 & 0.065 & 0.108 & 0.137 & 0.134 & 0.033 & 0.082 & 0.000 & 0.142\\
10 & 0.076 & 0.060 & 0.086 & 0.101 & 0.084 & 0.028 & 0.066 & 0.006 & 0.109\\
15 & 0.074 & 0.058 & 0.075 & 0.098 & 0.083 & 0.027 & 0.061 & 0.009 & 0.108\\
20 & 0.070 & 0.057 & 0.070 & 0.105 & 0.081 & 0.026 & 0.057 & 0.014 & 0.108\\
25 & 0.074 & 0.057 & 0.072 & 0.102 & 0.077 & 0.028 & 0.057 & 0.016 & 0.104\\
30 & 0.076 & 0.061 & 0.071 & 0.104 & 0.077 & 0.028 & 0.058 & 0.021 & 0.107\\
35 & 0.078 & 0.061 & 0.074 & 0.106 & 0.076 & 0.030 & 0.057 & 0.023 & 0.111\\
40 & 0.080 & 0.061 & 0.077 & 0.110 & 0.075 & 0.033 & 0.059 & 0.025 & 0.111\\
45 & 0.078 & 0.062 & 0.076 & 0.111 & 0.075 & 0.033 & 0.059 & 0.028 & 0.111\\
50 & 0.080 & 0.065 & 0.077 & 0.111 & 0.077 & 0.034 & 0.060 & 0.032 & 0.110\\
\hline
\end{tabular}
\caption{MAP values at each value of $N$ for the rankings using LDA for topic modelling.  Abbreviations: Topic (LDA order), Cit. (citation order), Bet. (betweenness), Cls (closeness), PR (PageRank), Dgr (degree), Inf. (influence), HS (hub score) and $\mbox{MC}_2$ is the Markov chain model of \cite{dwork2001rank}.}
\label{tab:ldaMap}
\end{table*}

\section{Conclusions} 

We proposed an approach for finding experts in a set of authors and their publications. The method uses well-known topic modelling algorithms LSI and LDA to identify authors within the query domain, and then construct a coauthorship graph using these authors. In turn, the graph is used for the extraction of expert rankings using a number of centrality measures. Furthermore, we explore the use of a rank aggregation approach to leverage the orderings and improve rankings. Computational results on the large Arnetminer dataset show that the citation ranking and betweenness in conjunction with LSI for topic modelling provide the most precise single-rank estimates of experts, however these rankings are improved significantly using aggregations. 

\section*{Acknowledgements} 
This work is funded by the Eurostars ERASM project.

\bibliographystyle{abbrv}
\bibliography{references}

\end{document}